\documentclass[a4paper, 12pt]{article}
\usepackage[utf8]{inputenc}
\usepackage{geometry}   
\usepackage{tabularx}
\usepackage{verbatimbox}
\usepackage{xcolor}
\usepackage{physics}
\usepackage{tabularx}
\usepackage{bbold}
\usepackage{amsmath}
\usepackage{amssymb}
\usepackage{natbib}
\usepackage{graphicx}
\usepackage{caption}
\usepackage{authblk}

\title{Causal nonseparability and its implications for spatiotemporal relations}
\date{  }
\author[1,2]{Laurie Letertre}
\affil[1]{Chair of Excellence in Philosophy of Quantum Physics, Institut N\'eel \& IPhiG, Universit\'e Grenoble Alpes, 38000 Grenoble, France}
\affil[1]{Institute of Philosophy, Czech Academy of Science, Prague, Czech Republic}

\begin{document}

\maketitle
\textbf{Declarations of interest}: none

\vspace{1cm}

\textbf{Acknowledgements}:

Many thanks to Vincent Lam and Cyril Branciard for their continuous encouragement and suggestions for the improvement of this paper. I am also thankful to Cristian Mariani, Claudio Calosi, Kerry Mckenzie and anonymous reviewers for their very helpful comments. This work was supported by the Agence Nationale de la Recherche under the programme ``Investissements d'avenir'' (ANR-15-IDEX-02) and the Formal Epistemology Project funded by the Czech Academy of Science.

\begin{abstract}
    Quantum nonseparability is a central feature of quantum mechanics, and raises important philosophical questions. Interestingly, a particular theoretical development of quantum mechanics, called the process matrix formalism (PMF), features another kind of nonseparability, called \textit{causal nonseparability}. The PMF appeals to the notion of quantum process, which is a generalisation of the concept of quantum state allowing to represent quantum-like correlations between quantum events over multiple parties without specifying a priori their spatiotemporal locations. Crucially, since the PMF makes no assumption about the global causal structure between quantum events, it allows for the existence of causally nonseparable quantum processes. Such processes are said to have an indefinite causal structure. This work aims at investigating the philosophical implications of causal nonseparability, especially for the notion of spatiotemporal relations. A preliminary discussion will first study the formal connection between quantum and causal nonseparability. It will be emphasised that, although quantum processes can be seen as a generalisation of density matrices, the conceptual distinction between the two notions yields significant differences between quantum and causal nonseparability. From there, it will be shown that, depending on the interpretative framework, causal nonseparability suggests some kind of indeterminacy of spatiotemporal relations. Namely, within a realist context, spatiotemporal relations can be epistemically or metaphysically indeterminate. Finally, it will be argued that, in spite of the disanalogies between standard and causal nonseparability, similar implications for spatial relations can already be defended in the context of standard quantum mechanics. This work highlights the potentially very fruitful explorations of the implications of quantum features on the conception of spacetime, keeping in mind that quantum and spacetime theories are expected to be unified in a future theory of quantum gravity. 
\end{abstract}



\section{Introduction}


Quantum entanglement is a key ingredient of quantum mechanics~\citep{schrodinger1935gegenwartige}, and might even be central for the very distinction between the quantum and classical theories~\citep{janotta2014generalized}. In a composite system made of several single sub-parts, this phenomenon refers to the situation for which it is impossible to attribute independent definite states to each of the sub-systems, and the composite system has to be considered as a whole; its sub-parts are said to be non-separable. Philosophically, quantum entanglement has important implications. First of all, it is at the core of the main conceptual puzzle arising from the development of non-classical physics, known as the measurement problem~\citep{maudlin1995three}. Secondly, entanglement allows for the observation of very peculiar, non-classical correlations between spacelike-separated events, which cannot be explained by any local causal model and are thus said to be nonlocal~\citep{bell1964einstein}.
Overall, many interpretations of quantum mechanics have been developed in order to solve the measurement problem and provide an account for nonlocality. Within a realist framework, which postulates a direct link between a theory’s ontology and that of the objective world, each interpretation commits to a particular set of assumptions regarding the properties of reality, by specifying the ontology of the theory and its dynamics. Which of those interpretations is the most successful is still an ongoing debate.

Yet, entanglement is not an exclusive property of standard quantum mechanics. Instead, it is also central in further theoretical developments of quantum physics, e.g. quantum field theory, and is expected to remain an important feature of quantum gravity (QG). Investigating the philosophical implications of entanglement in a broader theoretical context could shed a new light on the conceptual problems in quantum theory. Indeed, since (i) the way entanglement and nonlocality are accounted for is partly conditioned by our conception of spacetime, and (ii) a radical shift in our conception of spacetime is expected to take place as quantum physics is developed into more general theories (where gravity would be ultimately taken into account), looking at entanglement in new theoretical frameworks generalising standard quantum mechanics could help us understand the nature of entanglement and its connection to space and time in a radically new way (see section~\ref{back} for a development of these points). Identifying the relation between entanglement and spacetime may well be crucial for developing a consistent ontology of reality. 

Also in that spirit, there exists a particular theoretical development of quantum mechanics that features another kind of nonseparability, with an interesting connection with spatiotemporal notions.
This framework, called the process matrix formalism, makes no assumption about the global causal structure connecting quantum events. It allows for the existence of quantum processes (a generalisation of the concept of quantum state allowing to represent quantum-like correlations between events over multiple parties without specifying \textit{a priori} their spatio-temporal locations) that are \textit{causally non-separable}, for which one can see some analogy with the spatial non-separability involved in entangled systems. For a causally non-separable process, there is no definite causal order among its interacting parties. The corresponding causal structure is said to be indefinite. In operational terms, the probability distribution encapsulating the results of measurements performed on certain causally nonseparable processes may violate the causal equivalent of Bell inequalities, called causal inequalities. Such a distribution is said to be noncausal. 

The goal of this paper is, first, to discuss the connection between the notions of quantum and causal nonseparability. More precisely, since the notion of causal nonseparability is inspired by some analogy with that of quantum nonseparability, we will investigate the extent and the limits of this formal analogy. Secondly, in a realist framework, we will have a preliminary reflection regarding the potential implications of causal nonseparability for the notion of space(time). Regarding those two questions, two corresponding statements will be defended. First, while quantum processes can indeed be seen as generalisations of quantum states, the two notions are mathematically and, most importantly, conceptually distinct. This implies significant differences between quantum and causal nonseparability. Secondly, depending on the interpretative framework, causal nonseparability suggests some kind of indeterminacy of spatiotemporal relations. It is argued that, in spite of the disanalogies between standard and causal nonseparability, such implications for spatial relations can already be defended in the context of standard quantum mechanics. 

Before developing those questions (see sections~\ref{state1} and \ref{state2}), the next section will present the process matrix formalism and its central feature called causal nonseparability. 

\section{The process matrix formalism and causal nonseparability}
\label{basics}

        \subsection{Overview of the formalism}
         \label{mot}

The development of the process matrix formalism by \citet[]{oreshkov2012quantum} was motivated by the desire to provide a more general formalism for quantum mechanics in which no global predefined causal order is assumed between different quantum events, which are basically a pair of input and output physical systems\footnote{A distinction needs to be made between the \textit{classical} inputs (i.e. measurement settings) and outputs (i.e. measurements outcome) of a given quantum measurement, and the \textit{quantum} inputs (i.e. input quantum system) and outputs (i.e. output physical system) of a given quantum operation.} connected via some quantum operation\footnote{It is worth emphasising here that the notion of quantum event used in this work is therefore distinct from the notion of event conceived as a spacetime point.}. Within such a formalism, one can investigate whether more general causal structures than the definite (yet possibly dynamical) ones are compatible with quantum mechanics. 

\begin{figure}[ht]
  \centering
  \includegraphics[width=\textwidth]{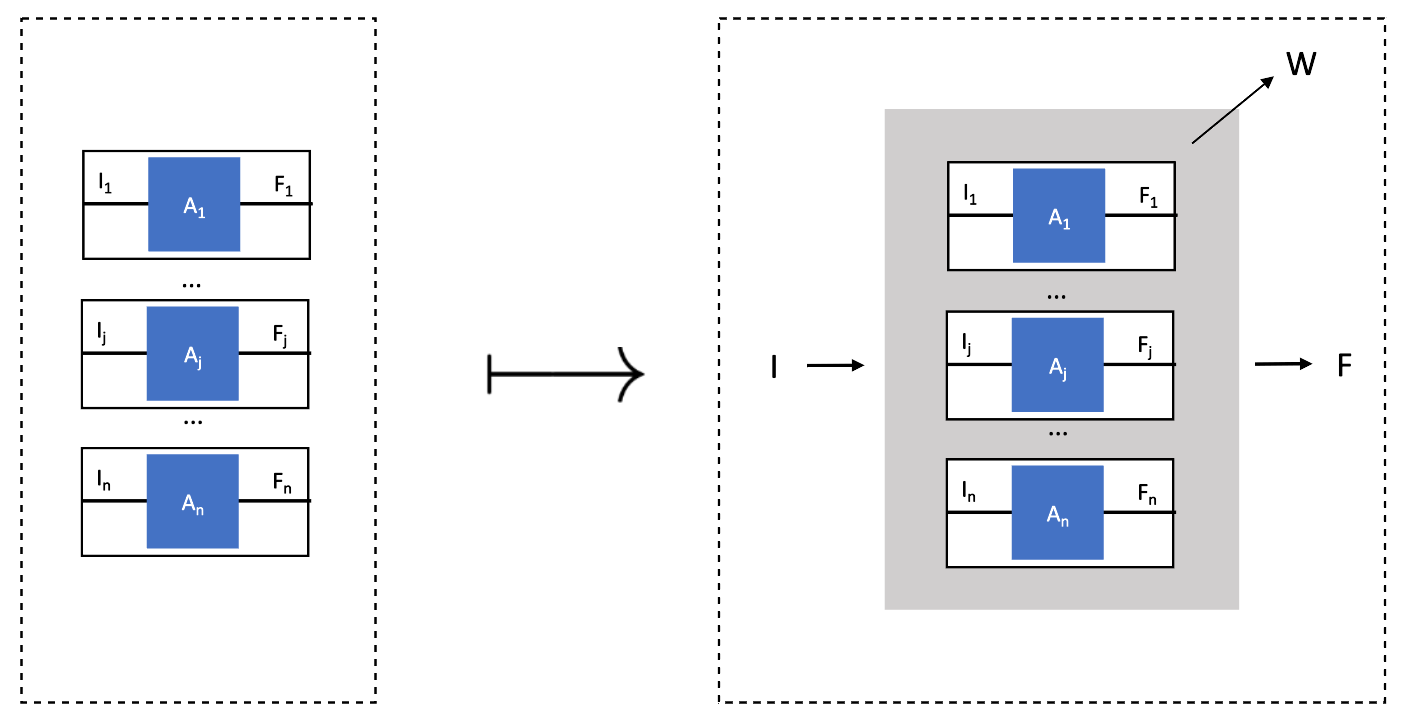}
  \caption{ }
  \label{process schema}
\end{figure}

The central object in the PMF is the \textit{process matrix} (denoted $W$), representing \textit{processes}, which are a list of joint probabilities for all possible local measurement's outcomes obtained in different isolated parties. In accordance with the very motivation of the formalism, it is postulated that the local experiments performed on quantum systems by different parties obey the rules of quantum mechanics, but no assumption is made regarding the spatio-temporal locations of these parties. Another way to understand what is a process matrix $W$ (representing a (quantum) process) is as a map applying $n$ local (quantum) operations (denoted $A_j$ with j going from 1 to $n$) over a global operation (see Fig.~\ref{process schema}, in which the labels $I$ and $F$ represent the input and output of the process, respectively, while $I_X$ and $F_X$ represent the input and output of the local operation $A_X$, respectively.)\footnote{The notion of quantum process is therefore distinct from other known notions of processes (e.g. as used in metaphysics).}. $W$ satisfies a set of conditions ensuring its consistency with valid probability distributions~\citep{oreshkov2012quantum}.

A formal discussion of process matrices will be developed in more detail in section~\ref{state1}.

         \subsection{Causal nonseparability}
         \label{def}
    
Before introducing the notion of causal nonseparability, it is useful to recall that of quantum nonseparability (or entanglement). In standard quantum mechanics, entanglement characterises certain quantum states of composite systems. Mathematically, a quantum state is represented either by a vector (denoted $\ket{\psi}$) in the Hilbert space assigned to the system, or, more generally, by a density matrix (denoted $\rho$) acting on that Hilbert space. A density matrix can encode either pure or mixed quantum states. The former are vectors in a Hilbert space, while the latter are probabilistic mixture of vectors.

Let two physical systems, denoted $1$ and $2$, be assigned a corresponding Hilbert space denoted $\mathcal{H}_1$ and $\mathcal{H}_2$, respectively. These systems form a composite system denoted $1-2$, of which the corresponding Hilbert space is the tensor product of $\mathcal{H}_1$ and $\mathcal{H}_2$. The quantum state of the composite system $1-2$ is \textit{separable}, or \textit{non-entangled}, if it can be formulated as follows: 

\begin{equation}
   \rho_{1-2} = \sum\limits_{i} q_i ~\rho^i_1 \otimes \rho^i_2
   \label{density}
\end{equation}
in which the index i sums over the classical probabilities ($q_i$) that subsystem x is in the (pure or mixed) quantum state $\rho^i_x$.   
    
In analogy with the definition of \textit{quantum nonseparability}, the notion of \textit{causal nonseparability} can be defined. While the former notion characterises \textit{quantum} relations among separate degrees of freedom, to which different Hilbert spaces are attached, by referring to quantum states of systems, the latter notion characterises \textit{causal}~\footnote{The specific way the word \textit{causal} is used will be clarified below.} relations among quantum events by referring to quantum processes. For those reasons, while we speak of \textit{quantum} separability of \textit{quantum states}, we speak of \textit{causal} separability of \textit{processes}.

Let there be two parties, Alice and Bob, performing local quantum operations ($A$ and $B$, respectively) on some quantum system. The way those local operations are combined to form a global structure is described by the bipartite process denoted $W^{A,B}$. By definition, $W^{A,B}$ is \textit{causally separable} if it can be decomposed as a probabilistic mixture of one-way (or no-) signalling causal processes \citep[]{oreshkov2012quantum, oreshkov2016causal}:

\begin{equation}
W^{A,B} = q W^{A\prec B} + (1-q) W^{B\prec A}   
\label{causalsep}
\end{equation}
where $q$ is a number between 0 and 1 and $W^{X\prec Y}$ represents a process for which the generated correlations, if they ever display signalling, are such that signalling takes place only from $X$ to $Y$ (excluding the possibility of both two-way signalling and signalling from $Y$ to $X$)~\footnote{It is important to emphasise that the notations $X \prec Y$ and $X \succ Y$ refer, in this paper, to the operational scenarios ``signalling is only possible from $X$ to $Y$'' and ``signalling is only possible from $Y$ to $X$'', respectively. Those relations, of which the relata are \textit{quantum} events, are referred to as ``causal relations'', where the adjective ``causal'' takes on a specific operational meaning. Indeed, the relation $X \prec Y$ describes a situation where the inputs of party $X$ can influence the outputs of party $Y$. Such a relation can therefore be understood causally from an interventionist perspective (I am thankful to an anonymous reviewer for this suggestion): $X \prec Y$ if and only if an intervention (here, the choice of classical inputs) in party $X$ can change what happens (here, the classical output) in party $Y$. This intervention should respect the following requirements: after the intervention (i.e. the choice of input) is made, $X$ shouldn't be influenced by any other variables; any influence of the intervention over party $Y$ goes through party $X$; and the intervention should be statistically independent of any other variables influencing $Y$ that are not in a chain of influences going through $X$~\citep[Chap.~3]{woodward2005making}. Within the process matrix formalism, those requirements are satisfied as the parties operate in closed laboratories and the free choice of the measurement settings is assumed.}. Although we will focus exclusively on the bipartite case here, it is worth mentioning that a generalisation of Eq.~\eqref{causalsep} for multipartite processes has been developed in \citet{oreshkov2016causal} and \citet{wechs2018definition}.

From Eq~\eqref{causalsep}, we see that a causally separable process is a convex combination of processes compatible with a given, fixed causal structure. We say that a causally separable process is therefore a process that is compatible with an underlying \textit{definite} causal structure\footnote{More precisely, the global causal structure among the quantum events described by that particular process is definite. Of course, a causally separable process (therefore being a probabilistic mixture of processes with a fixed causal structure) can itself be part of a wider causally nonseparable process, which would be incompatible with a definite causal structure. Hence, the origin of the probabilistic mixture can be either a classical ignorance regarding the actual global structure of the process (referred to as a ``proper mixture''), or an indefinite causal structure at a broader scale (referred to as an ``improper mixture''). This situation is similar to that of quantum nonseparability~\citep[p.~110]{nielsen2002quantum}.}. That is, the causal ordering between the quantum events is definite, even when an imperfect preparation procedure only yields a given causal order with a certain probability.

A quantum process generates specific correlations depending on the experiment that is performed. Let's consider a joint measurement performed by two observers, Alice and Bob, with a given set of classical inputs $x$ and $y$ corresponding to Alice’s and Bob’s measurement setting, respectively\footnote{It is important to emphasise here that the parties involved in a given process need not be conscious agents. The classical inputs of parties can be free variables (i.e. not correlated with properties in their causal past or in the rest of the experimental setup) of the systems involved. For this reason, the adjective “operational” used in the context of the process matrix formalism may take on a less literal signification than when referring to operations performed by conscious agents. However, this does not undermine the causal understanding of the structure encoded in a process matrix suggested in footnote 5. Indeed, the interventionist account holds no matter whether agents are present or not~\citep[p.94, 103, 104]{woodward2005making}.}. The corresponding joint probability to obtain the classical outcomes $a$ for Alice and $b$ for Bob is denoted $P^{AB}(a,b|x,y)$. If one focuses on those correlations instead of on the process itself, it is possible to provide an operational characterisation of the corresponding causal structure featured by the process. A correlation $P^{AB}(a,b|x,y)$ is said to be \textit{noncausal} if it can't be expressed as a probabilistic mixture of probabilistic distributions $P^{A\prec B}(a,b|x,y)$ and $P^{B\prec A}(a,b|x,y)$, which are valid probability distributions compatible with a fixed causal order between $A$ and $B$~\citep[]{oreshkov2012quantum, branciard2015simplest}. The causal order $A \prec B$ means, operationally, that $B$ cannot signal to $A$, and reciprocally for the causal order $B \prec A$. It is possible to derive algebraic inequalities that are violated by noncausal correlation~\citep{oreshkov2012quantum, branciard2015simplest}. Those inequalities are called \textit{causal inequalities}.

Although causal nonseparability is necessary for noncausal correlations \citep{oreshkov2012quantum, wechs2018definition}, previous work showed that a causally non-separable process will not necessarily generate correlations that will violate a causal inequality \citep[]{oreshkov2016causal, araujo2015witnessing}. Its non-sufficiency can in particular be demonstrated by the example of the quantum switch (see section~\ref{QS}), which is causally nonseparable but does not lead to any noncausal correlations. So far, no physical protocol that generates noncausal correlations has yet been found~\citep{wechs2021quantum, PhysRevLett.127.110402}\footnote{Noncausal correlations have been observed in the literature, but those do not correspond strictly speaking to the notion presented in this work. \citet{ho2018violation} showed that it was possible to observe noncausal correlations by relying on a specific protocol in which the parties perform operations within a spatially localised laboratory, but within an extended interval of time. Such laboratories are therefore not strictly closed, and causal cycles become allowed between the parties. Other works showed that one could obtain noncausal correlations by use of post-selection~\citep{oreshkov2016operational, silva2017connecting, araujo2017quantum, milz2018entanglement}.}\footnote{The failure (so far) to develop a protocol violating a causal inequality means that noncausal correlations might more likely exist in another physical realm (such as that of quantum gravity) or not at all. 

The existence of noncausal correlations would “merely” ensure a model-independent confirmation that a specific physical phenomenon is taking place in some causally nonseparable processes. Yet, it is conceptually possible to consider that causal nonseparability does refer to a (new) physical feature of the world, even without the presence of noncausal correlations. Indeed, noncausal correlations imply the observations of two-way signalling, which is a much stronger constraint than the mere physicality of causal nonseparability, which implies, e.g., a computational advantage~\citep{chiribella2012perfect, colnaghi2012quantum, araujo2014computational, facchini2015quantum, feix2015quantum, guerin2016exponential, ebler2018enhanced, salek2018quantum, chiribella2021indefinite, mukhopadhyay2018superposition, procopio2019communication} not relying on the existence of two-way signalling. The physicality of causal nonseparability does not depend on that of noncausality. 
As a result, the discussion of realist approaches towards causal nonseparability is not fully dependent on the existence of noncausal correlations, and one can still speak of “genuine indefinite causal orders” generating causal correlations.

Similarly, the existence or nonexistence of noncausality does not influence the preferred philosophical reading of indefinite causal orders themselves, in the same way that the existence of nonlocal correlations does not imply a preferred interpretation of quantum entanglement. In the present case, causal nonseparability (when considered as pointing at an objective physical phenomenon) has the potential to bring novel interpretational features (such as those developed later in this paper) without having these features necessarily implying the existence of noncausal correlations as a consequence of their presence. As a result, the preference for epistemic or metaphysical readings of ICO depends rather on potential theoretical virtues or philosophical preferences rather than being constrained by empirical evidence (such as noncausality).
}. 

    \subsection{A famous example of a causally nonseparable quantum process: the quantum switch}
    \label{QS}
    
\begin{figure}[ht]
  \centering
  \includegraphics[width=\textwidth]{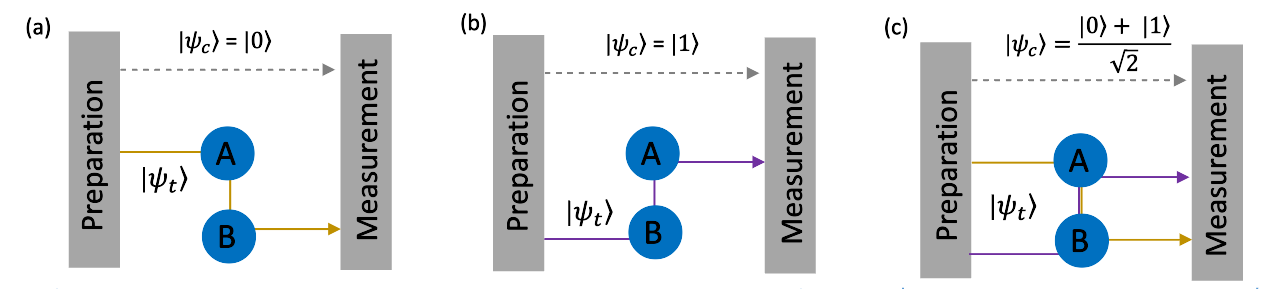}
  \caption{}
  \label{QSTOTAL}
\end{figure}

\begin{figure}[ht]
  \centering
  \includegraphics[width=\textwidth]{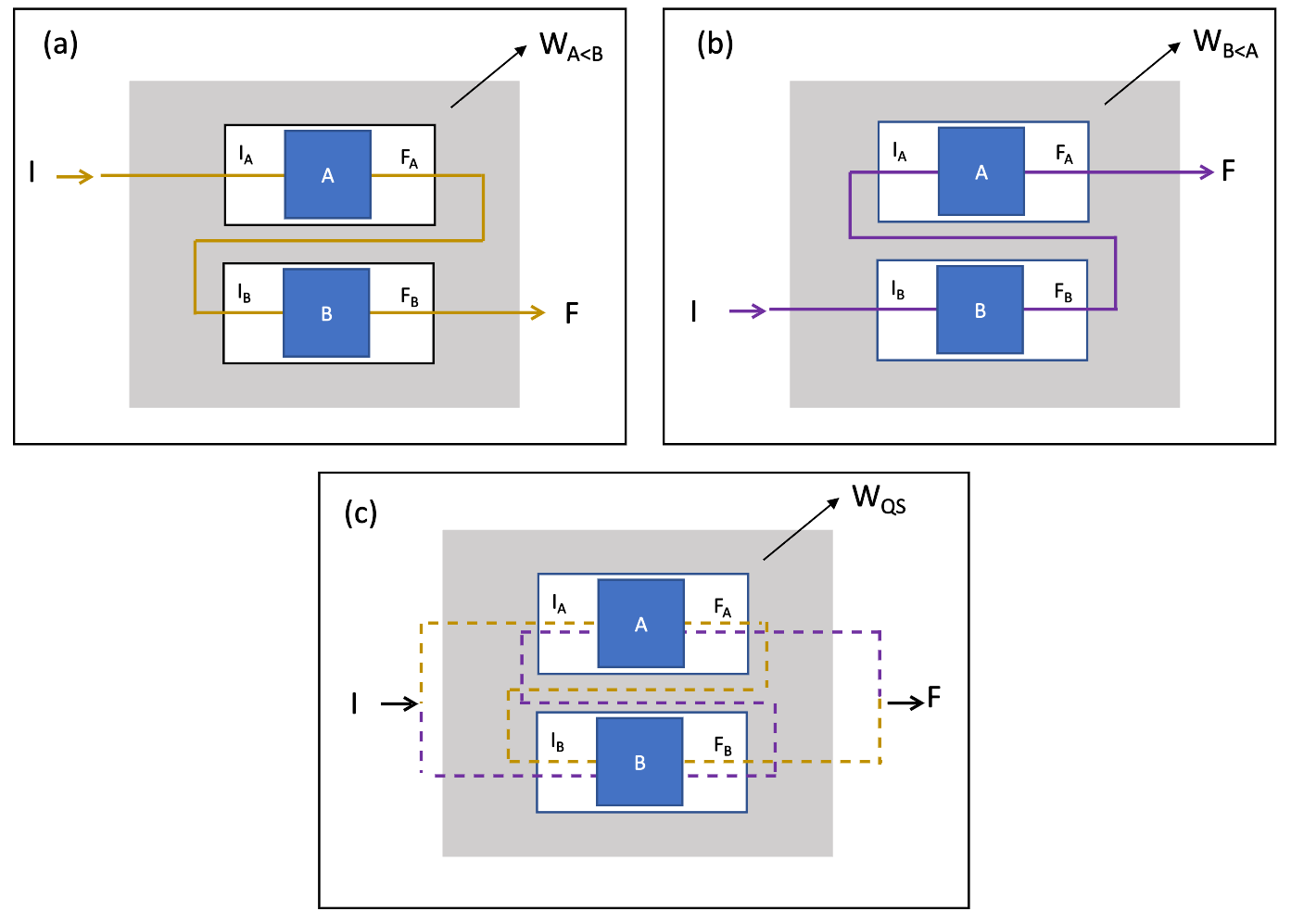}
  \caption{}
  \label{QS simple}
\end{figure}

\citet{chiribella2013quantum} have imagined a circuit, called the quantum switch (QS), of which the process matrix has been proved to be causally nonseparable \citep[]{oreshkov2016causal, araujo2015witnessing}. In this process, two local operations ($A$ and $B$) are mapped over a global operation. The two operations $A$ and $B$ are performed by isolated parties Alice and Bob, respectively. They will both manipulate a shared physical system called the \textit{target system}. Which of Alice and Bob will receive the target system first depends on the state of an additional qubit, called the \textit{control system}. If the control qubit is in the state $\ket{0}$, Alice will perform her operation on the target system before it is sent to Bob (see Fig.~\ref{QSTOTAL} (a)), and reciprocally if the control qubit is in the state $\ket{1}$ (see Fig.~\ref{QSTOTAL} (b)). A third party, Fiona\footnote{This third party, Fiona, is actually crucial for the QS to be causally nonseparable. Since Fiona ($F$) receives the control qubit after the target system has left both laboratories $A$ and $B$, and performs some operation on it, the quantum switch is strictly speaking a tripartite process.  Tracing out its process matrix over the third party would lead to an improper mixture of fixed causal orders, i.e. a causally separable process matrix. However, it can be shown that a tripartite process $W$ in which one party has no outcome system for his/her operation (we throw it away) is causally separable if and only if it can be expressed as $W~=~q~W^{A \prec B \prec F} ~+~(1-q)~W^{B \prec A \prec F}$. Hence, as long as Fiona has no output system for her operation (it is the case within the quantum switch), we find ourselves in a situation in which the causal order is indeterminate among two operations. For that reason, the tripartite causal nonseparability of the quantum switch amounts, to a certain extent, to a bipartite case of causal nonseparability.} (corresponding to the measurement procedures in Figures~\ref{QSTOTAL}(a), (b) and (c)), will operate on the control qubit after Alice and Bob both acted on the target system, so that the information about the causal order between Alice and Bob will be erased. In the quantum switch, the initial state of the control qubit is in a superposition of the states $\ket{0}$ and $\ket{1}$. The causal order between operations $A$ and $B$ becomes then entangled with the control qubit's state (see Fig.~\ref{QSTOTAL} (c)). 
However, it is common to use a slight misuse of language to describe that situation, and say that the QS is in a superposition of causal orders between Alice and Bob\footnote{As indicated earlier, the parties do not necessarily involve conscious agents. As a result, the philosophical interpretation of causal nonseparability is not impacted by the debate regarding whether conscious agents can be in a quantum superposition. I am thankful to an anonymous referee for having raised that issue.}. Indeed, if we simplify the quantum switch to consider only Alice and Bob acting on the target system, there are two possible fixed causal structures linking them\footnote{The process matrix formalism presupposes that each of the operations forming the set that is mapped over some global operation by the quantum process is performed once and only once. An extension of the formalism allowing multiple rounds of information exchange for each party has been developed in \citep{hoffreumon2021multi}.}: either only operation $A$ can influence $B$ (denoted $A \prec B$) (see Fig.~\ref{QS simple} (a)), or reciprocally (denoted $B \prec A$) (see Fig.~\ref{QS simple} (b)). In the truncated description of the QS, the global structure combining operations $A$ and $B$ is in a superposition of these definite structures $A \prec B$ and $B \prec A$ (see Fig.~\ref{QS simple} (c)).
The quantum switch is such that the operations cannot be said to be performed in a definite causal order. We say that the underlying causal structure is \textit{indefinite}, and displays an \textit{indefinite causal order} (ICO).

   \subsection{On the physicality of the quantum switch}
        \label{open}

Distinguishing the process matrices referring to physical quantum processes from process matrices that are just mathematical artefacts of the formalism, without any reference in the world itself, is a difficult task that is currently still under investigation (see e.g. \citep{araujo2017purification, wechs2021quantum})\footnote{The question of the physical status of causally nonseparable processes has pragmatic implications, since causal nonseparability leads to computational advantages when implemented in circuits performing certain tasks~\citep{chiribella2012perfect, colnaghi2012quantum, araujo2014computational, facchini2015quantum, feix2015quantum, guerin2016exponential, ebler2018enhanced, salek2018quantum, chiribella2021indefinite, mukhopadhyay2018superposition, procopio2019communication}. Yet, because the validity of such implementations is still debated, the question remains open regarding whether such a computational advantage is genuine or simulated.}. Yet, there are at least certain causally nonseparable processes that have a rather intuitive physical implementation~\citep{wechs2021quantum}, namely the quantum switch (see section~\ref{QS}) 
which has been physically implemented in different ways~\citep{procopio2015experimental, rubino2017experimental, goswami2018indefinite, wei2019experimental, guo2020experimental}. 

Yet, the idea that those implementations are physical realisations of a proper indefinite causal order has been criticised in the literature. The process matrix formalism assumes that each party is perfectly isolated from the rest of the world, and performs its operation once and only once. The reason is that quantum processes need to satisfy these features in order to be valid processes\footnote{A valid process matrix satisfies certain constraints in order to ensure that only valid probability distributions (i.e. non-negative and normalised) are generated when applying the generalised Born rule. Those constraints are detailed in \citep{oreshkov2012quantum}.}. Moreover, the violation of these criteria would weaken the meaningfulness of the concept of indefinite causal order as a purely quantum phenomenon. As an example, in a bipartite scenario in which parties are not isolated, the possible causal relations between $A$ and $B$ would not be exhausted anymore by the set \{``$A \prec B$'', ``$B \prec A$''\}. Indeed, one could imagine that $A$ and $B$ influence each other, forming a causal cycle. Such a configuration could not be expressed as a probabilistic mixture of definite causal orders ``$A \prec B$'' and ``$B \prec A$'', yet would have a classical understanding~\citep{maclean2017quantum}. 

For these reasons, it is therefore important for any implementation of the quantum switch to ensure that each party is well isolated, and that its operation is performed once and only once. \citet{oreshkov2019time} showed that local operations in a particular class of processes (including the quantum switch) are assigned time-delocalized Hilbert spaces with respect to which the causal structure is definite and involves causal cycles. This provides a clear mathematical argument for saying that the assumption that local operations are performed once and only once can be verified in practice through quantum process tomography.

Another objection to existing implementations of the quantum switch claims that a genuine implementation of an indefinite causal order would be only achievable by a \textit{gravitational} quantum switch (i.e. involving an actual superposition of spacetime metrics) ~\citep{zych2019bell,paunkovic2020causal}. It is argued that the quantum switch cannot provide a correct description of spacetime at quantum scales, because it assumes a classical non-relativistic spacetime. This objection arises upon translating ICOs to spatiotemporal relations, which is a move discussed in section~\ref{P2.4}. The present work is sympathetic to such an objection, and the attitude that is defended here is that indefinite causal orders can be seen as pointing towards an existing tension between certain quantum features and classical spacetime. A (metaphysical) investigation of ICOs might therefore help develop useful conceptual tools that might be relevant in more fundamental quantum theories of spacetime (see the end of section~\ref{metaICO}).

From the above discussions, it seems then interesting and meaningful to consider, or at least assume, that indefinite causal orders can be found in physically implementable processes. The question now will be to elucidate what it could correspond to in the world (section~\ref{state2}). Since that question will be discussed under the assumption of scientific realism, section~\ref{Re} will discuss the motivations of such a stance in the context of quantum foundations. Yet, before entering those discussions, we will first explore the formal analogy existing between quantum and causal nonseparability, in order to emphasise the analogies and disanalogies between these concepts. This step will ensure a good formal understanding before moving on to the interpretational questions.

\section{About the formal analogy between quantum and causal nonseparability}
\label{state1}

We will now discuss the formal connection between the notions of quantum and causal nonseparability. Because those two notions are based on the density matrix and the process matrix, respectively, we will start by contrasting these two mathematical objects.

\begin{figure}[ht]
  \centering
  \includegraphics[width=\textwidth]{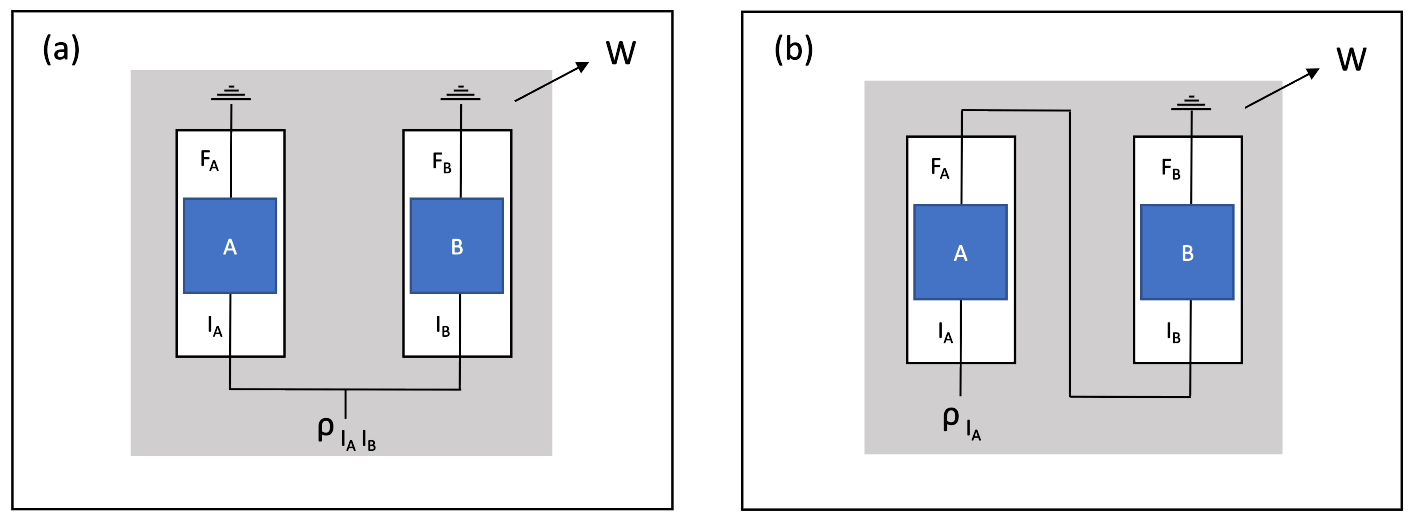}
  \caption{}
  \label{processgen}
\end{figure}

The process matrix $W$ can be seen as a generalisation of the density matrix $\rho$~\citep{oreshkov2012quantum}. The sense in which $W$ generalises $\rho$ is that $W$ can reduce to a density matrix in specific cases, e.g. when two systems (initially described by the joint input quantum state $\rho_{I_A I_B}$) undergo distinct operations ($A$ and $B$, respectively) after which the output systems are thrown away (see Fig.~\ref{processgen} (a)). Yet, more generally, a process matrix can encode information that cannot be expressed by density matrices, e.g. when one and the same system (initially described by the quantum state $\rho_{I_A}$) undergoes two successive operations ($A$ and $B$), while the final output system is thrown away (see Fig.~\ref{processgen} (b)).

In spite of this idea that the process matrix generalises, in some sense, the density matrix, those two concepts are two distinct mathematical objects of a different nature. Indeed, these two objects describe different notions: 

\begin{itemize}
    \item The \textbf{density matrix} describes the \textit{quantum state} of a given system.
    \item The \textbf{process matrix} describes the \textit{process that provides a mathematical representation of certain relations between} quantum events (i.e. a pair of quantum input and output connected by a quantum operation). 
\end{itemize}

While a quantum state is, mathematically, a vector in a Hilbert space or a density matrix acting on that space, a quantum operation is a map describing how such vectors or density matrices are transformed into other vectors or matrices. 
At this stage, it is useful to insist on an important point. While the process matrix can be seen, in the specific sense presented above, as a generalisation of the density matrix, and allows thereby to encode transformations across time, we see that it \textit{does not} mean that process matrices are generalisations of density matrices in the sense of relating quantum states at different spatial \textit{and temporal} locations. Process matrices allow representing relations among quantum events, independently of (i) the systems involved, and (ii) of the quantum operations. The first claim is true when the preparation procedure for the input system of the process is not fixed by the process itself, and is rather left as a quantum event corresponding to an additional party. In that case, the process does not encode a particular quantum state, hence, a particular system. The second claim is true because, while the process describes how the operations' inputs and outputs are connected together, the operations themselves are not specified. A process matrix is therefore profoundly different than a mathematical tool merely connecting quantum states at possibly different times and locations. The temporal dimension enters the picture because we shift from a picture in which the objects of interest are the quantum states to a picture in which the objects of interest are quantum operations. 

As a conclusion of this comparison, we see clearly that the process matrix $W$ and the density matrix $\rho$ are distinct mathematical objects, with different inner structures. A process matrix $W$ does not represent a quantum state, but a \textit{process} that \textit{relates} different quantum events involving physical systems. 
The process matrix generalises the concept of density matrix in that it can, in certain cases, reduce to (possibly composite) quantum states. Yet, it also allows representing transformations thereof. More generally, process matrices describe how local quantum operations are combined to form a single global operation.
To sum up, process matrices shift the focus from relations between quantum states to relations between quantum events, and involve both spatial and temporal dimensions. They describe the global structure (possibly indefinite) underlying different quantum events.
Hence, by asking what are the kind of underlying causal structures compatible with valid process matrices, one investigates the causal structures possibly compatible with quantum mechanics.

Based on the above distinction between density and process matrix, we can now emphasise the formal difference between quantum and causal nonseparability. The former expresses that for some composite systems, the global quantum state is nonseparable, the quantum states of the sub-systems being indefinite. The latter expresses that for some processes, the corresponding process matrix is nonseparable, and the order among the quantum events within that process is indefinite. Hence, the two kinds of nonseparability are conceptually very different, as they describe quantum correlations among very different notions, namely quantum states in the case of quantum nonseparability and quantum events in the case of causal nonseparability. 

\section{About the realist interpretation of causal nonseparability}
\label{state2}

Having clarified the notions of process matrices and causal nonseparability, and how they differ from the notions of density matrices and quantum nonseparability, this section will investigate the kind of implications that one would face when adopting a realist attitude towards the process matrix formalism.

    \subsection{Scientific realism and the process matrix formalism}
    \label{Re}

First, a word on the realist framework guiding this discussion. Since all currently known causally nonseparable processes can be described in standard quantum mechanics\footnote{In the sense that the quantum states of the systems involved (as well as their evolution) can be described using standard quantum mechanics.}~\citep{abbott2020communication}, it would be a perfectly coherent attitude to look at their characterisation within the process matrix formalism (hence, at process matrices and causal nonseparability) as purely formal, i.e. not capturing any novel objective features of the world. Yet, it remains an open question whether there exist causally non-separable processes in nature, that are not describable within standard quantum mechanics. Since the scenario in which causal nonseparability points towards novel objective features of the world may give rise to new physical and metaphysical discussions, it is the one developed further in this section. 

A realist account of the process matrix formalism agrees with the view that its central object, the process matrix, refers to some objective features of the world. In that context, one has to articulate the exact meaning of a process matrix and the new idea that, for a causally nonseparable process, there is no well-defined causal structure among the quantum events related by the process. 

This task is vast and can be undertaken in a variety of ways. In this paper, we will adopt a broader viewpoint for exploring the general consequences of causal nonseparability, which does not commit to a particular ontology for the theory or a particular solution to the measurement problem. Then, we will motivate a natural connection to the idea of indefinite spatiotemporal structures. The role of spacetime in non-relativistic quantum mechanics is to provide a fixed background stage with a Galilean geometry for events to take place. This status still holds within the process matrix formalism\footnote{See \citet{paunkovic2020causal} for a discussion of that particular point.}. Yet, in spite of spacetime's supposedly passive role in this theory, we will see that adopting a realist attitude towards causal non-separability can have philosophical implications for spatiotemporal relations, as the notion of indeterminate spatiotemporal relations suggests. 
From there, we will explore different readings of that indeterminacy (namely the epistemic and metaphysical ones), and the way they might impact the notion of spacetime itself. It is worth noting that although one particular reading of indeterminacy will not be compatible with all quantum ontologies and solutions to the measurement problem, the reflections presented below can be developed independently of such considerations. 

    \subsection{The quantum switch as a case study for indefinite causal orders}
    \label{QS-pres}

Let's consider the particular causally nonseparable quantum process presented in section~\ref{QS}, called the quantum switch. At this stage, causal nonseparability and indefinite causal orders are purely formal concepts. Assigning them a meaning is the next step. 
The indefiniteness of causal orders in the QS is easily identified as due to the entanglement of the causal order between Alice and Bob with the state of an additional system. Notice that this description implicitly considers the notion of causal order as theoretically equivalent to some physical observable. This is technically relevant, as it is possible to define the operators $\mathcal{O_{AB}} = \ket{0}\bra{0}$ and $\mathcal{O_{BA}} = \ket{1}\bra{1}$ (with $\ket{0}$ and $\ket{1}$ being the control qubit's states from Fig.~\ref{QSTOTAL} (a) and (b)) acting on the process \textit{vector} of the QS (denoted $\ket{W_{QS}}$)\footnote{Which can be expressed as $\ket{W_{QS}} = \dfrac{1}{\sqrt{2}} (\ket{\psi}^{A^I} \ket{\mathbb{1}}^{A^O B^I} \ket{\mathbb{1}}^{B^O F^t} \ket{0}^{F^c} + \ket{\psi}^{B^I} \ket{\mathbb{1}}^{B^O A^I} \ket{\mathbb{1}}^{A^O F^t} \ket{1}^{F^c})$,  with $\ket{\psi}^{A^I}$ ($\ket{\psi}^{B^I}$) being the quantum state of the target system as input of operation A (B), $\ket{0}^{F^c}$ and $\ket{1}^{F^c}$ are the quantum states of the control system as input of the third party (Fiona), and $\ket{\mathbb{1}}^{X Y}$ is the Choi vector of the identity channel sending the output quantum state of operation $X$ on the input of operation $Y$, i.e. it is the transformation of the identity channel by the use of the (pure) Choi isomorphism~\citep[]{choi1975completely}. The Hilbert space to which each vector belongs is specified as a superscript to avoid any ambiguity. The Hilbert spaces denoted $X^I$ and $X^O$ correspond to the input and output Hilbert spaces of operation X, respectively, and those denoted $F^c$ and $F^t$ correspond to Fiona's input for the control qubit and target system, respectively.}, and of which the eigenvectors would correspond to the processes defined in Fig.~\ref{QSTOTAL} (a) and (b), respectively. Hence, the causal order between operations $A$ and $B$ can be considered as an observable property of $\ket{W_{QS}}$ with values ``$A \prec B$'' and ``$B \prec A$''. In the quantum switch, this quantum property and the state of the control qubit are the relata of an entanglement relation. 

Assigning a meaning to this entanglement relation can be done either by considering a full-fledged account of quantum mechanics, or by adopting a specific reading of entanglement without entering the full details of a specific solution to the measurement problem. We will rely on the latter option, as it allows us to keep a more focused attention to the meaning of entanglement itself without taking into account other constraints not relevant to this specific discussion. 
From that perspective, it is enough to acknowledge that, in standard quantum mechanics, entanglement can be read as an \textit{objective} feature of nature, or as an \textit{effective} phenomenon reflecting an imperfect knowledge of the observer. As we will see below, the meaning of the indefinite causal order in the QS (resulting from its entanglement with the state of an additional system) can be discussed along those two possible routes: either the indeterminacy of causal orders is understood as objective (i.e. metaphysical), or epistemic. 

We will not engage, however, in a deeper discussion of indefinite \textit{causal relations}. Instead, we will now shift from the notion of causal structures to that of spatiotemporal ones. This will allow us to explore the implications of ICO on space and time. This is attractive as space and time \textit{unambiguously}, in contrast to causation, constitute a central concept in the ontology of the world. Additionally, this move will allow us to get closer to the exploration of a pressing question in fundamental physics (at least within our scientific realist framework), namely the fate of spacetime in a quantum world. As discussed in section~\ref{Re}, spacetime is classical in standard quantum mechanics. Yet, this status is expected to change in a more general theory of quantum gravity. It is interesting to investigate whether there already exist tensions between this assumption of a classical spacetime and certain quantum features of non-relativistic, non-gravitational quantum theory.

\subsection{From causal to spacetime structures}
\label{P2.4}

Causation and spacetime are intimately connected in various theories. For example, in Lewis' standard account of supervenience, causation is described as emerging from the mosaic of actual facts (localised in space and time). By contrast, certain approaches to quantum gravity describe relativistic spacetime as emerging, to some extent, from more fundamental causal structures~\citep[ch. 2]{OON}.
For the purpose of this argument, we will focus on the way causal relations and spacetime geometry constrain each other in the theory of relativity. 
Indeed, in the context of general relativity, it has been shown that for spacetime manifolds that are past and future distinguishing, the geometry of spacetime is determined by its causal structure up to a conformal factor~\citep{hawking1976new, malament1977class}. In other words, there is an isomorphism between the spacetime structure (the metric) and the causal structure for spacetime manifolds that possess certain properties allowing us to consider them as physical\footnote{``Physical'' spacetimes possess indeed certain properties (such as, e.g., being past and future distinguishing or globally hyperbolic) that correspond to specific constraints on their conformal structure. See \citep{minguzzi2008causal} for a review. Some words of caution are in order here, as \citet{earman1972notes} pointed out that there exist ``exotic'' spacetime structures not obeying the ``past and future distinguishing'' constraint, but still licensed by general relativity, and there is no reason to discard such spacetime structures as valid candidates for modelling our spacetime.}. Here, the notion of causal structure (hereafter named ``relativistic causal structure'') is a mathematically defined structure describing the type of causal relation (or lack thereof) that can obtain between any pair of spacetime points, in agreement with the requirement that causes of a given effect are located in the past lightcone of that event.

We see from those considerations that drawing a connection between the concepts of causality (via the mathematically defined ``relativistic causal structure'') and spatiotemporality is rather natural in the context of relativity. Connecting the causal structures encoded in quantum processes to spatiotemporal relations is not as straightforward. The reason is that causal orders within process matrices are operationally defined, and independent, by construction, from spatiotemporal notions. A given process matrix encodes the relations among inputs and outputs of various \textit{quantum} events, the network of such relations being called the ``global causal structure'' (hereafter named ``operational causal structure'') of the process under consideration. By contrast, relativistic causal structures describe how \textit{spacetime} events are connected as constrained by relativity. 

Yet, a connection between the ``operational causal structure'' and the spacetime manifold can be defended. Since the quantum events that are connected in the operational causal structure of a process necessarily take place in spacetime, they must correspond to some (definite or indefinite) spatiotemporal locations. The causal influences possibly taking place between these spacetime regions, as constrained by the theory of relativity, are encoded in the ``relativistic causal structure''. As such, there must be some connection between the operational and relativistic causal structures. \citet{vilasini2022embedding} have closely investigated this question\footnote{\citet{vilasini2022embedding} use the term ``information-theoretic-event'' instead of that of quantum event. It can easily be verified that the two terms refer to the same notion.}. 
The authors proved a no-go theorem stating that the implementation in a (definite) relativistic spacetime of any process with an indefinite causal order necessary implies that the quantum systems involved are delocalised in spacetime. One sees that the indeterminacy of the operational causal structure gets translated into indefinite spatiotemporal locations in a definite and relativistic spacetime. 

Now, one needs to assign a meaning to these indefinite spatiotemporal locations. One option is to see this delocalisation as purely epistemic, i.e. corresponding to an effective description of the world, instead of referring directly to the objective spatiotemporal properties of the system. Those would be well defined, as it would be the case, e.g., in a Bohmian reading of the situation. In that case, relativity would be rejected, which is why we would escape the no-go theorem of \citet{vilasini2022embedding}. Another option is to read indefinite spatiotemporal locations as metaphysically indeterminate. Yet, it would be misleading to consider that the spatiotemporal properties \textit{of the systems involved} are indefinite. Indeed, as we saw in section~\ref{state1}, indefinite causal order is really an instance of indeterminacy of the order between quantum operations, and not of the states of systems. Hence, indefinite spatiotemporal locations would rather be an instance of \textit{spacetime} indeterminacy. This reading amounts to establishing a correspondence between a description of delocalised systems in a definite spacetime structure and that of localised systems in an indefinite spacetime. Such a formal correspondence has been explicitly shown in the context of the gravitational quantum switch in \citep{dimic2020simulating} (see section~\ref{metaICO} for a discussion of the gravitational quantum switch).

To summarise, it follows from these joint results that an indefinite operational causal order encoded in a quantum process can be seen as corresponding to \textit{either} an effective (i.e. epistemic) description of spatiotemporally delocalised systems in a definite spacetime background, \textit{or} an objective (i.e. metaphysical) description of spatiotemporally localised systems in an indefinite spacetime background. Both these options will be discussed in the next section, with an emphasis on the one bearing metaphysical implications for spacetime. 

\subsection{Indefinite spatiotemporal relations}
\label{metaICO}

Let's now discuss the meaning behind indefinite spatiotemporal relations according to the two above-mentioned options. Similarly to what was explained in section~\ref{QS-pres}, the indeterminacy of a spatiotemporal structure can have an epistemic or a metaphysical (i.e. ontic) origin. An epistemic indeterminacy of spatiotemporal relations would originate from an imperfect \textit{knowledge} of what the objective spatiotemporal relations actually are. The notion of spatiotemporal relation would be semantically adequate, so that it is, in principle, capable of referring to the objective spatiotemporal features of the world. However, sometimes (e.g. in the QS), instead of referring directly to those features, it refers to the (imperfect) \textit{knowledge} we have of such matters\footnote{This ignorance being not classical, in the sense that the global situation cannot be described by a probabilistic mixture of corresponding possible spatiotemporal structures.}. On the other hand, a metaphysical indeterminacy of spatiotemporal relations would not come from a lack of knowledge regarding the objective spatiotemporal relations, but rather from objective features of the world. In other words, the objective spatiotemporal relations would themselves be metaphysically indeterminate. 

In order to illustrate the first option, namely the epistemic reading of indefinite spatiotemporal relations, it is useful to specify the proposal in more detail. Indeed, the kind of ontology and dynamics that can be associated to shape a full-fledged account of the process matrix formalism constrain the nature of the indeterminacy. An epistemic kind of indeterminacy would be present in realist accounts of the process matrix formalism where relevant theoretical elements (here, the process matrix associated to the experimental setting under consideration) would be considered as not referring to objective aspects of the world, while the universal process\footnote{The universal quantum process would describe the relations existing between all the quantum events of the universe.} would\footnote{As a recall, we consider that being a realist towards the process matrix formalism means that the process matrix refers to objective features of nature. The above-mentioned attitude towards the process matrix, in which only the universal process is objective, is analogous to the way the wavefunction is understood in some ``$\psi$-ontic'' realist approaches to quantum mechanics. For instance, in Bohmian mechanics, the universal wavefunction refers to objective aspects of the world, while the wavefunction of individual systems does not.}. 

An analogy with particular realist approaches towards standard quantum mechanics can be drawn: quantum indeterminacy has an epistemic reading when, e.g., entanglement is read as some effective phenomenon. As an example, an account of quantum mechanics can assign a nomological status to the universal wavefunction~\citep{durr1995bohmian, esfeld2014ontology}, while the quantum states assigned to sub-systems of the universe are mere mathematical tools that do not refer to something in the world. 

The counterpart of that strategy, in the context of the PMF, would be to read the universal process matrix as a law of nature\footnote{There is already a debate regarding whether the universal wavefunction, or the universal density matrix, are acceptable candidates as laws of nature. As suggested by \citet[]{chen2020quantum}, a law of nature is reasonably expected to be simple, fixed by the theory, generating motion and not referring to things in the ontology of the world. The universal quantum process might violate some of these criteria, e.g. as it does not generate motion. Whether this invalidates a possible nomological status for the universal process matrix is left as an open question at this stage. For example, one could be satisfied with laws of nature imposing global constraints on the ontology rather than ruling a proper dynamics.}. The corresponding constraint would then have a holistic nature. As a result, a quantum process taking place in a sub-region of the universe would provide an approximate description of what happens in that region: it would ignore the holistic character of the world's constraints and represent what happens in that sub-region \textit{as if} it were isolated from the rest of the world. We could then say that quantum processes assigned to sub-regions of the universe wouldn't refer to objective features of the world. To sum up, the universal process matrix would be objective, but the processes associated to any particular experimental setting would not be real. Instead, they would describe merely effective phenomena. 

When assuming the physicality of causal nonseparability, taking the indefiniteness of causal orders as epistemic might seem to be an easy fix for explaining the phenomenon. Yet, in all generality, an epistemic account of ICO must still provide an explanation for the way the non-actual causal orders seem to have a physical influence within the global process due to the presence of quantum interferences between different causal orders. Indeed, it can be shown that the process matrix of the quantum switch is expressed as the following sum of terms: 

\begin{equation}
    W_{QS} = 1/2 (W^{A \prec B} + W^{B \prec A} + \ket{W}^{A \prec B} \bra{W}^{B \prec A} + \ket{W}^{B \prec A} \bra{W}^{A \prec B})
    \label{interference}
\end{equation}
in which $W^{X \prec Y} = \ket{W}^{X \prec Y} \bra{W}^{X \prec Y}$, with $\ket{W}^{X \prec Y}$ representing a process vector with the causal structure ``$X \prec Y$''. The two first terms of Eq.~\eqref{interference}, of the form $W^{X \prec Y}$, are process matrices having the definite causal structure ``$X \prec Y$'', while the two last terms of the form $\ket{W}^{X \prec Y} \bra{W}^{Y \prec X}$ correspond to interference terms. It is their very presence that makes $W_{QS}$ causally nonseparable by preventing its formulation as a probabilistic mixture of processes with a definite causal structure. In the same way that accounts of standard quantum mechanics in which quantum nonseparability reflects an incomplete description of reality (e.g. Bohmian mechanics) still need to account for the presence of quantum interference, an epistemic account of ICO needs to explain the underlying physical reason for its presence, despite the fact that the actual causal structure is definite~\footnote{Otherwise, it might be said that an epistemic reading of ICO would lack a certain explanatory appeal regarding the presence of indeterminate causal orders, since the origin of the \textit{in principle} lack of knowledge about causal relations would ultimately remain as a brute posit about causation.}. Whether a given epistemic account of ICOs would be convincing and preferable to a metaphysical reading would depend on the details of the account, and is left to the appreciation of individuals. 

In contrast to the previous case, the scenario of a metaphysical indeterminacy for ICOs would indicate that the spatiotemporal relations themselves (and not our knowledge of them) are sometimes indefinite.
Again, this proposal can be exemplified by specifying the proposal in more detail. Metaphysical indeterminacy can be obtained within accounts of the PMF in which the process matrix of a given causally nonseparable process would be taken as referring to some objective content of the world. 

An analogy can be drawn with the situation in standard quantum mechanics. In that context, one can see the universal wavefunction as a real entity located at the fundamental level of the world's ontology. This view is often called $\psi$-realism~\citep{albert2013wave}. A similar approach in the context of the PMF would be to see the universal process matrix as a real and fundamental entity living in a high-dimensional Hilbert space\footnote{This space would be the tensor product of the Hilbert spaces assigned to the quantum input and output of all the quantum events of the universe.} which would correspond to the real fundamental space in which the world's fundamental ontology is located. Spacetime would then be emergent from that fundamental space, and quantum processes describing what happens in a given sub-region of our familiar spacetime could refer to objective, yet derivative\footnote{There is a debate regarding whether metaphysical indeterminacy affecting derivative entities is eliminable (see, e.g., \citep{glick2017}).}, entities. As another potential analogy\footnote{I am thankful to an anonymous reviewer for having suggested exploring this approach in the context of the process matrix formalism.}, one can look at Rovelli's relational interpretation. The interpretation would involve an ontology containing a kind of “relational” entities (of which the properties are encoded in the wavefunction of the corresponding physical systems). Their nature needs to be articulated, e.g. with metaphysical tools such as (moderate) ontic structural realism~\citep{candiotto2017reality, oldofredi2021bundle}.
It seems that applying a similar strategy to the process matrix formalism can coincide with a metaphysical reading of the indefinite causal orders. Causally nonseparable process matrices would encode a certain property (causal orders) that is supervenient to, or on a par with (depending on the kind of ontic structural realism one considers), the relation of causal nonseparability. The quantum switch can then arguably be understood in terms of the determinable-based account of metaphysical indeterminacy~\citep{wilson2013determinable}: a determinable (``having a definite causal order'') carried by the process has more than one determinate (``$A \prec B$'' or ``$B \prec A$'') in a relativised fashion (i.e. relatively to a specific state of the control's qubit). One could say that the causal order in the quantum switch is metaphysically indefinite \textit{in virtue of} being supervenient to, or \textit{on a par} with, the relation of causal nonseparability\footnote{I am grateful to Claudio Calosi for having suggested, in a different context, this sort of possible connection between metaphysical indeterminacy and ontic structural realism.}.

%
A more general approach to a metaphysical reading of ICOs can be discussed. The idea of metaphysical indeterminacy within realist accounts of quantum mechanics has been discussed in the literature (almost) independently of any solution to the measurement problem (see~\citep{mariani2021ind} for a review). The idea is to take at face value the lack of value definiteness for certain physical observables arising from different sources such as quantum superpositions, entanglement or non-commutativity of observables. As discussed in section~\ref{QS-pres}, it is technically relevant to interpret causal nonseparability as a source of lack of value definiteness for the QS's property ``being such as to allow for a one way signalling between $A$ and $B$ (or no signalling at all)''. Upon shifting from operational causal structures to spatiotemporal ones (as constrained by relativity), this property becomes ``Having operations $A$ and $B$ being temporally ordered (or being spacelike separated)''.
From there (modulo an adequate background ontology), the indeterminacy of the property's value can be taken as metaphysical and articulated according to different accounts. 
According to a \textit{supervaluationist} view of metaphysical indeterminacy~\citep{barnes2011theory}, spatio-temporal relations are metaphysically indefinite when the truth-value of propositions involving those relations is indeterminate. Indeterminacy would then be a primitive, and constitutes a brute, unanalysable fact of nature. In the case of the QS, this would be expressed as follows: it is indeterminate which of the two states of affairs obtains among ``operation $A$ is in the relativistic past of operation $B$ (or spacelike separated from $B$)'' and ``operation $B$ is in the relativistic past of operation $A$ (or spacelike separated from $A$)''.
Alternatively, and as advocated in influential accounts of quantum indeterminacy, a \textit{determinable-based} account of metaphysical indeterminacy \citep{wilson2013determinable} would say that spatiotemporal relations are indefinite when a spatiotemporal determinable property fails to have a unique determinate, i.e. it either simultaneously presents more than one determinate, or lacks any determinate at all. 
The metrical properties of spacetime can be analysed \textit{prima facie} as candidate determinables that would lack a unique determinate in the QS. The exact signification of the metric will depend on the specific philosophical account of spacetime in the context of Galilean relativity (namely the substantivalist and relationalist accounts). In the former case, spatiotemporal relations are fundamental and independent of the matter distribution. The indeterminate metric would then correspond roughly to an indeterminate spacetime ``container'' for the content of the whole universe. In the latter case, the metric is viewed as the ensemble of geometrical relations existing among bodies, and hence, supervenes on matter. An indeterminate metric in that scenario would correspond to indefinite geometrical relations among matter. 

To sum up, the metaphysical reading of ICOs, while not formulated in the most fundamental context, still allows pointing at some tension existing between the idea of a classical spacetime background and quantum features such as nonseparability, as it leads to non-classical consequences for spacetime itself. This idea of an indeterminate spacetime metric is however arguably better discussed within a more fundamental context. It is indeed interesting to note, at this stage, some possible connections with existing work in the field of quantum gravity, in which spacetime's nature and characteristics can undergo tremendous changes. 

The notion of superposed gravitational fields is allowed in mainstream approaches to quantum gravity~\citep{paunkovic2020causal}, which keeps the discussion very general. As discussed in section~\ref{open}, \citet{zych2019bell} have proposed a thought experiment, called the gravitational quantum switch, based on the basic principles of quantum mechanics and general relativity (namely the quantum superposition principle and gravitational time dilation\footnote{In general relativity, time dilation refers to a difference in the measured temporal duration between two events as measured by two different clocks located at different spatiotemporal locations between which there exists a difference in gravitational potential~\citep{einstein2016relativity}.}). The thought experiment shows that those principles lead to a superposition of spacetime regions, which amounts to the presence of indefiniteness of causal orders. More precisely, indefinite causal orders originate from the entanglement of temporal orders between time-like events. While this thought experiment shares some similarities with the indefinite causal orders discussed in this work (the quantum switch displays entanglement of the causal order between two quantum events with a control system's state), \citet{zych2019bell} argue that there is a fundamental difference between the two settings. Indeed, the indefinite causal orders described by the process matrix formalism are embedded in a classical spacetime, and only two specific quantum events are described by a non-classical order as the result of a causal order entangled with some additional system. On the contrary, in the thought experiment involving gravitational effects, it is a whole spacetime region that is concerned with entangled temporal orders.
Indefinite temporal orders are the result of a superposition of a mass distribution, itself (as prescribed by general relativity) linked to spacetime geometry. That way, an explicit superposition of spacetime itself is obtained~\citep{paunkovic2020causal}. It is in that context only that the authors would qualify spacetime as \textit{non-classical}. 

This thought experiment reflects the importance to explore the implications of indefinite causal orders within the context of a theory of quantum gravity. Yet, this comparison allows us to put forward an argument in favour of the interpretation of causal nonseparability as an instance of metaphysically indeterminate spatiotemporal relations. First of all, the shift (in the context of the process matrix formalism) from indefinite causal order to indefinite spatiotemporal relations allows us to discuss causal nonseparability in terms of indefinite spacetime, which is a notion that is straightforwardly present, at the formal level, in most of the approaches to quantum gravity. Second, in the face of the multiplicity of ways in which this indeterminacy of spacetime can be interpreted, the metaphysical reading has the advantage of allowing for a form of metaphysical continuity across different quantum theories. Indeed, metaphysical indeterminacy, when applied to both contexts\footnote{See \citep{cinti2021lack} for a discussion of metaphysical indeterminacy of spacetime in the context of quantum gravity.}, would characterise the same entities, namely spatiotemporal relations. In the face of the threat to scientific realism that represents the potential lack of metaphysical progress as physics evolves (see \citep{mckenzie2019curse}), metaphysical indeterminacy of spatiotemporal relations ensures a metaphysical continuity across deeply different scientific theories. 
    
\section{Back to standard quantum mechanics}
\label{back}

Interestingly, the idea that quantum physics suggests particular implications for \textit{space} can already be found in the context of standard quantum mechanics. Indeed, quantum states encoding the spatial position of some physical system can be indefinite, i.e. in a superposition of eigenstates of the position observable. One can also encounter entangled quantum states encoding the values of the position observable. For example, in the two-slit experiment, the position of the electrons can be seen as being entangled with the state of the slits (i.e. open or closed). The system [slits + electron] is then quantum nonseparable, and the state of the electron (describing its (observable) position) is indefinite.

Again, the nature of this indefiniteness (epistemic or metaphysical) needs to be specified within a more detailed account. While a Bohmian approach would consider that the spatial locations of a quantum system are always well defined (i.e. the indeterminacy is epistemic), embracing quantum indeterminacy would consider that spatial locations (and possibly spacetime itself) are metaphysically indeterminate. 

It is worth noting that a notion of indefinite temporal relation would be complicated to explore in the context of standard quantum mechanics, since the very notion of a time observable in quantum mechanics is nontrivial\footnote{See \citet{butterfield2013time} for a review of the different roles that time can take on within a theory, namely a coordinate of spacetime (i.e. an external independent variable, or parameter) versus a function of other quantities of the system \textit{in} spacetime (i.e. a dynamical variable). The literature on time as a physical variable is vast (see, e.g., \citep{giovannetti2015quantum, erker2017autonomous, brunetti2010time, hilgevoord2002time}). While time is widely used as an external parameter in quantum mechanics, its measurement as a (definite or indefinite) variable remains non-trivial.}~\footnote{Relatedly, earlier work suggested the development of the idea of \textit{temporal} entanglement in standard quantum mechanics, but the task encountered technical difficulties. See \citep{glick2019timelike, adlam2018spooky} for a review.}. The process matrix formalism, by allowing to access the notion of causal order (hence, in some indirect sense according to section~\ref{P2.4}, that of temporal ordering), provides a way of investigating to what extent temporal relations can display quantum properties. 

To conclude, it seems that, provided we relate quantum events to spacetime events (which is necessary to connect indefinite causal orders in causally nonseparable processes to indefinite spatiotemporal relations), the (formal) indefiniteness of \textit{spatial relations} that can be encountered in the context of quantum nonseparability of quantum states can be extended to the (formal) indefiniteness of \textit{spatiotemporal relations} encountered in the context of causal nonseparability of process matrices.

\section{Conclusion}

It was argued in this paper that a realist attitude towards causal non-separability can, upon certain hypotheses, lead to the metaphysical indeterminacy of spatiotemporal relations. 

After discussing the formal definition of causal nonseparability, we relied on the quantum switch as a case study for exploring indefinite causal orders within a scientific realist framework. Upon shifting from a notion of causal structure to a notion of spatiotemporal structure, indefinite causal orders translated into indefinite spatiotemporal relations. It was argued that there exists an interesting argument for adopting a metaphysical stance towards those indefinite spatiotemporal relations, namely that it ensures a form of metaphysical continuity across different quantum theories. This highlights the idea that causal nonseparability, albeit being defined in a non-fundamental setting, can already point towards a tension between a classical notion of spacetime and certain quantum features.

It was finally highlighted that important consequences for spatial relations can already be defended in standard quantum mechanics. Hence, in spite of the disanalogies between quantum and causal nonseparability, both concepts can support substantial implications for (the properties of) spatio-(temporal) relations. 
Such results highlight the potentially very fruitful explorations of the implications of quantum features on the conception of spacetime, keeping in mind that quantum and spacetime theories are expected to be unified in a future theory of quantum gravity\footnote{See \citep{lam2022quantum} for a related discussion.}.

\bibliographystyle{agsm}
\bibliography{biblio2}
\end{document}